\documentclass[aps,prb,reprint,showpacs,amsmath,amssymb,amsfonts,floatfix]{revtex4-1}

\usepackage{bm,graphicx,epstopdf,color}
\usepackage[bookmarks,bookmarksopen,bookmarksnumbered,colorlinks,linkcolor=red,linktocpage,citecolor=blue,urlcolor=cyan,
pdfpagemode=UseOutline]{hyperref}

\def\bea{\begin{eqnarray}}
\def\eea{\end{eqnarray}}
\def\ben{\begin{equation}}
\def\een{\end{equation}}
\def\benu{\begin{enumerate}}
\def\enu{\end{enumerate}}

\def\bei{\begin{itemize}}
\def\eei{\end{itemize}}
\def\benu{\begin{enumerate}}
\def\enu{\end{enumerate}}

\def\sss{\scriptscriptstyle\rm}

\def\1var{(\bx_1...\bx\N)}

\def\br{{\bf r}}

\def\bx{{x}}

\def\x{_{\sss X}}

\def\xc{_{\sss XC}}

\def\Hxc{_{\sss HXC}}

\def\N{_{\sss N}}

\def\sph_int{ {\int d^3 r}}

\newcommand{\intd}{\mathrm{d}}
\newcommand{\vect}[1]{\mathbf{#1}}
\newcommand{\matelem}[3]{\left\langle #1 \left| #2 \right| #3 \right\rangle}

\providecommand{\abs}[1]{\left|#1\right|}

\newcommand{\parref}[1]{(\ref{#1})}

\newcommand{\xcs}{{\sss XC}}

\begin{document}
\title{Direct calculation of exciton binding energies with time-dependent density-functional theory}
\author{Zeng-hui Yang and Carsten A. Ullrich}
\affiliation{Department of Physics and Astronomy, University of Missouri, Columbia, MO 65211, USA}
\date{\today}

\begin{abstract}
Excitons are electron-hole pairs appearing below the band gap
in insulators and semiconductors. They are vital to
photovoltaics, but are hard to obtain with time-dependent
density-functional theory (TDDFT), since most standard
exchange-correlation (xc) functionals lack the proper
long-range behavior. Furthermore, optical spectra of bulk
solids calculated with TDDFT often lack the required resolution
to distinguish discrete, weakly bound excitons from the continuum.
We adapt the Casida equation formalism for molecular
excitations to periodic solids, which allows us to obtain
exciton binding energies directly. We calculate exciton binding energies
for both small- and large-gap semiconductors and insulators,
study the recently proposed bootstrap xc kernel [S. Sharma {\em
et al.}, Phys. Rev. Lett. {\bf 107}, 186401 (2011)], and extend
the formalism to triplet excitons.
\end{abstract}

\maketitle

\section{Introduction}
\label{sect:intro} Excitons arise from electron-hole attraction in
gapped periodic systems such as bulk insulators and semiconductors, as well as in many types of nanoscale systems,
polymers and biomolecules.\cite{Koch2006,Scholes2006} Bound excitons appear in optical
spectra of extended systems as discrete absorption peaks below the quasiparticle
gap, while continuum excitons enhance the band-edge
absorption.\cite{HK09} Excitons play an important role in
photovoltaics, where photo-excited excitons propagate to
heterojunctions and dissociate to yield currents. Although the
phenomenological Wannier model\cite{W37,D56,HK09} describes
excitons qualitatively well, it is not quantitatively suitable to be used in
real applications where {\em ab initio} computation is required.

The most important characteristic of bound excitons is their
binding energy, defined as the difference between the
quasiparticle gap and the excitation frequency of the exciton.
The Bethe-Salpeter equation (BSE), a many-body method, is
the standard way of calculating exciton binding energies in
periodic systems,\cite{ORR02} due to its accuracy. However, the scaling of the computational cost for BSE
versus system size is not favorable, and the use of the BSE has
therefore been limited to moderate system sizes, despite recent
progress.\cite{Puschnig2002,Fuchs2008,Ramos2008,Setten2011}

Thanks to the balance of accuracy and computational cost,
density-functional theory (DFT) and time-dependent
density-functional theory (TDDFT) are popular \textit{ab
initio} methods for electronic structure and dynamics.\cite{MMNG12,U12} Instead of approaching the many-body
problem directly, density-functional methods construct a
noninteracting Kohn-Sham system with the same electronic
density as the original interacting system, which is much
easier to solve than the original many-body problem. Despite
some additional difficulties for periodic systems (in
particular, the severely underestimated gap), TDDFT methods are
gaining popularity in solid-state
physics.\cite{ORR02,Botti2007,Yabana2012}

TDDFT is a formally exact theory for electron dynamics, but in
practice the exchange-correlation (xc) kernel must be
approximated. It has been notoriously difficult to get excitons
in
TDDFT:\cite{Reining2002,Kim2002,Sottile2003,Marini2003,Nazarov2011}
local and semilocal xc kernels that work well in finite systems
do not yield bound excitons in solids, since they lack a long-range
part.\cite{GGG97} The recently proposed long-range correction
(LRC) xc kernel\cite{GGG97,BSVO04,Bruneval2006} allows bound
excitons to be obtained from TDDFT, but empirical input is
required.

Aside from the difficulty to find good xc kernels, there is
another problem. TDDFT approaches for periodic systems
typically calculate the optical spectrum via the dielectric
function; but exciton binding energies in semiconductors are
usually in the meV range, which means that bound excitons
require a high frequency resolution to be distinguished from
the continuum. This makes the calculation numerically
 demanding. One could increase the frequency resolution
near the region of interest, but this requires knowing the
exciton binding energies beforehand. As a consequence, most
existing TDDFT studies of excitons are either for materials
with  strongly bound excitons far away from the band edge such
as $\mathrm{LiF}$ or $\mathrm{Ar}$,\cite{Sottile2007} or describe the enhancement of the band-edge continuum spectrum
due to excitonic effects.\cite{ORR02,Botti2007}

We recently proposed an alternative TDDFT approach for obtaining
excitonic binding energies directly.\cite{TU08,TLU09,YLU12} The
approach was applied to one-dimensional model
systems,\cite{YLU12} where we showed that TDDFT within the
adiabatic approximation can yield more than one exciton if
local-field effects are included. We also considered several
bulk solids and found that TDDFT, using xc kernels with
appropriate long-range behavior, can yield excitonic binding
energies in the right range.\cite{TLU09} However, this earlier
work remained somewhat inconclusive due to several
simplifications (most notably, a two-band approximation, a
rather small $\vect{k}$-space grid, and a real-space representation of
the xc kernel which resulted in a loss of accuracy).

In this paper we present a systematic computational study of
the lowest excitonic binding energies in common zincblende and
wurtzite semiconductors as well as in large-gap insulators. We
extend our earlier work\cite{TLU09} in several ways: we go
beyond the two-band approximation and include, in principle, an
arbitrary number of bands; this can be viewed as the
solid-state analog of the Casida approach for molecular
excitation energies. \cite{C96} Furthermore, we extend the
formalism to include triplet excitons, and we test the
so-called bootstrap xc kernel.\cite{SDSG11}

Atomic units ($e=\hbar=m_e=1/(4\pi\epsilon_0)=1$) are used
throughout this paper unless mentioned otherwise.

\section{Theoretical background}
\label{sect:theory}

In semiconductors, the binding between an electron and a hole
is usually weak: such electron-hole pairs are designated as
Wannier excitons. The Wannier model\cite{W37,D56,HK09}
describes such excitons in analogy to positronium, where the
effect of the material environment is introduced by the
effective mass and the dielectric constant. The Wannier
equation for excitons is given by
\begin{equation}
\left[-\frac{\nabla^2}{2m_r}-\frac{1}{\epsilon r}\right]\psi_\nu(\br)=E_\nu\psi_\nu(\br),
\label{eqn:theory:Wannier}
\end{equation}
where $m_r=(m_h^{-1}-m_e^{-1})^{-1}$ is the reduced effective
mass, $\epsilon$ is the dielectric constant of the material,
and $E_\nu$ and $\psi_\nu$ are excitonic binding energy and
wave function, respectively. The binding energy is the most
important property for excitons, defined as the difference
between the quasiparticle gap and the excitonic excitation
energy. Despite the simplicity of Eq.
\parref{eqn:theory:Wannier}, its exciton binding energies can be
fairly accurate for common semiconductors such as GaAs
\cite{Ulbrich1985} and $\mathrm{Cu_2O}$,\cite{Uihlein1981}
since the model can be derived as an approximation to the BSE
many-body theory\cite{SR66} (assuming the effective Bohr radii
of excitons are much greater than the lattice constant).

Excitons in large-gap materials (such as LiF and Ar) are
strongly bound and localized within a single crystal unit. The
Bohr radii of these so-called Frenkel excitons are small, so
the Wannier model does not describe Frenkel excitons well.
However, from the point of view of an \textit{ab initio}
electronic structure theory, there is no conceptual difference
between Wannier and Frenkel excitons, as they all are just
excitations of the many-body system. The collective character
of the excitons distinguishes them from other excitations,
i.e., they arise from superpositions of many single-particle
excitations.

Aside from their collective quasiparticle character, excitons are normal
optical excitations. TDDFT has been successful in treating
excitations in finite systems, and its use for periodic systems
is increasing. TDDFT solves a non-interacting time-dependent
system described by the time-dependent Kohn-Sham equation:
\begin{eqnarray}
i\frac{\partial}{\partial t}\phi(\br,t)
&=&
\bigg[-\frac{\nabla^2}{2}+v_\text{ext}(\br,t)+v_\text{H}(\br,t)
\nonumber\\
&&
{}+v\xc(\br,t)\bigg]\phi(\br,t)\:, \label{eqn:theory:TDDFT}
\end{eqnarray}
where $v_\text{ext}$ and $v_\text{H}$ are the external
potential and the Hartree potential, respectively, $v\xc$ is
the exchange-correlation (xc) potential, and $\phi$ is a time-dependent
Kohn-Sham orbital. $v\xc$ is defined as the one-body
multiplicative potential with which the solution of Eq.
\parref{eqn:theory:TDDFT} reproduces the density of the
interacting system, and it is the only part that needs to be
approximated in practice. One can obtain information about the
excitations in the interacting system by propagating Eq.
\parref{eqn:theory:TDDFT} under an external perturbative
potential,\cite{Yabana2006} but the more convenient approach is
to work in the frequency domain from the beginning.

The linear response function\cite{GiulianiVignale} $\chi=\delta
n/\delta v_\text{ext}$ describes the first-order density change caused by a
change in the external potential, and thus
determines the optical spectrum. In TDDFT,\cite{Gross1985}
the response function in reciprocal space is obtained, in principle exactly, as
\begin{eqnarray}
\chi_{\vect{G}\vect{G}'}(\vect{q},\omega)&=&\sum_{\vect{G}''}\bigg[\delta_{\vect{G}_1\vect{G}_2}-
\sum_{\vect{G}_3}\chi_{s,\vect{G}_1\vect{G}_3}(\vect{q},\omega) \nonumber\\
&\times&
f_{\text{Hxc},\vect{G}_3\vect{G}_2}(\vect{q},\omega)\bigg]^{-1}_{\vect{G}\vect{G}''}
\chi_{s,\vect{G}''\vect{G}'}(\vect{q},\omega),
\label{eqn:theory:dysonlike}
\end{eqnarray}
where $\chi_s$ is the linear response function of the Kohn-Sham
system,\cite{U12} $f\Hxc=f_\text{H}+f\xc$ with the Hartree
kernel $f_\text{H}=\delta v_\text{H}/\delta n$ (in reciprocal
space, $f_{\rm H} =
4\pi\delta_{\vect{G}\vect{G}'}/|\vect{q}+\vect{G}|^2$), and the
xc kernel $f\xc=\delta v\xc/\delta n$. All quantities in Eq.
\parref{eqn:theory:dysonlike} are matrices in reciprocal
space, where $\vect{q}$ belongs to the first Brillouin zone,
and the $\vect{G}$'s are reciprocal lattice vectors.

The optical absorption spectrum of a periodic system is described by the
macroscopic dielectric function
$\epsilon_\text{M}$.\cite{ORR02} One calculates $\epsilon_\text{M}$ from $\chi$ by
\begin{equation}
\epsilon_\text{M}(\omega)=\lim_{\vect{q}\to0}\frac{1}{1+4\pi\chi_{00}(\vect{q},\omega)/q^2}.
\end{equation}
Calculating $\epsilon_{\rm M}$ on a frequency grid yields the spectrum.

The so-called head ($\vect{G}=\vect{G}'=0$) of the xc kernel gives the
largest contribution to the change from $\chi_s$ to $\chi$, and
the contributions from bigger $\vect{G}$'s decay rapidly. Thus
the sums in Eq. \parref{eqn:theory:dysonlike} can be restricted
to a small number of reciprocal lattice vectors, which reduces
the computational effort significantly.

One can select the frequency $\omega$ in Eq.
\parref{eqn:theory:dysonlike}, so the calculation can be focused
on the region of interest instead of the entire spectrum. It
might appear that in Eq. \parref{eqn:theory:dysonlike} one can
choose any frequency resolution, but it is implicitly limited
by the number of Kohn-Sham excitations included in $\chi_s$. A
part of the spectrum is directly obtained from this approach,
but there is no way of knowing which Kohn-Sham excitations
contribute to a specific peak. Considering the continuum nature
of the spectra of periodic systems, these details are of course
rarely needed.  For excitons, however, the binding energies are
not explicitly given in this approach, unlike in the much simpler
Wannier model. Due to the discrete nature of bound excitons,
the Kohn-Sham excitation composition is useful for
interpretations; but this information is not available in this
approach.

In finite systems the low-lying excitations are discrete, and
there is a more efficient way to obtain them than scanning the
frequency range with Eq. \parref{eqn:theory:dysonlike}. The
idea is to describe electronic excitations as eigenmodes of the
system.\cite{Ullrich2009} The Casida equation\cite{C96} then transforms the TDDFT
linear-response equation into the transition space spanned by
single-particle Kohn-Sham excitations:
\begin{equation}
\left(\begin{array}{cc}\mathbf{A} & \mathbf{B}\\ \mathbf{B} & \mathbf{A}\end{array}\right)
\left(\begin{array}{c}\mathrm{X}\\ \mathrm{Y}\end{array}\right)=
\omega\left(\begin{array}{cc}-\mathbf{1} & \mathbf{0}\\ \mathbf{0} & \mathbf{1}\end{array}\right)
\left(\begin{array}{c}\mathrm{X}\\ \mathrm{Y}\end{array}\right),
\label{eqn:theory:Casida}
\end{equation}
where the matrix elements of $\mathbf{A}$ and $\mathbf{B}$ are
\begin{equation}
\begin{split}
A^{(ij)(mn)}(\omega)&=(\epsilon_j-\epsilon_i)\delta_{im}\delta_{jn}+F\Hxc^{(ij)(mn)}(\omega),\\
B^{(ji)(mn)}(\omega)&=F\Hxc^{(ji)(mn)}(\omega),
\end{split}
\end{equation}
with the Hartree-exchange-correlation (Hxc) matrix $F\Hxc$ for spin-unpolarized systems
\begin{eqnarray}
F^{(ij)(mn)}\Hxc(\omega)
&=&
2\int\intd^3r\int\intd^3r'\;\phi_i(\br)\phi_j^*(\br)f\Hxc(\br,\br',\omega)
\nonumber\\
&& \times
\phi_m^*(\br')\phi_n(\br'). \label{eqn:theory:FHxc}
\end{eqnarray}
The factor of 2 in $F\Hxc$ accounts for the spin and the $\phi$'s
are the ground-state Kohn-Sham orbitals. The indices $i$, $j$,
$m$, $n$ all represent full sets of quantum numbers, where $i$,
$m$ denote occupied orbitals and $j$, $n$ denote unoccupied
orbitals.

Most of the currently available xc kernels are frequency
independent, in which case Eq.
\parref{eqn:theory:Casida} becomes an eigenvalue problem. The
explicit matrix formulation of Eq. \parref{eqn:theory:Casida}
is suitable for discrete excitations in finite systems. The
excitation frequencies of the system are explicitly given by
the eigenvalues $\omega$. The eigenvector $\mathrm{X}$
together with $\mathrm{Y}$ describes how the Kohn-Sham
 excitations combine to form the excitation in
the real system. The optical spectrum can be calculated with
$\mathrm{X}$ and $\mathrm{Y}$.\cite{ORR02} The widely used
Tamm-Dancoff approximation (TDA) sets the matrix $\mathbf{B}$ to zero
and hence neglects the correlation between excitations and
de-excitations. We have shown that the TDA can often be a better
choice for excitons than the exact calculation with Eq.
\parref{eqn:theory:Casida},\cite{YLU12} and we therefore employ the TDA
throughout this paper.

The difficulty of obtaining excitons in TDDFT mainly comes from
the requirement on the xc kernel for periodic systems, that it
needs a $q^{-2}$ behavior in the $\vect{q}\to0$
limit.\cite{GGG97} This long-range behavior is necessary in
order to produce non-zero head ($\vect{G}=\vect{G}'=0$) and
wing ($\vect{G}=0$ or $\vect{G}'=0$) contribution to $\chi$
in Eq.
\parref{eqn:theory:dysonlike} and to $F\xc$ in Eq.
\parref{eqn:theory:FHxc}. In periodic systems these contributions dominate over
the so-called local-field effects [contributions from  the body
($\vect{G}\ne0$ and $\vect{G}'\ne0$) of $f\xc$].
 Common local and semi-local xc kernels, such as the
adiabatic local-density approximation (ALDA), lack this
long-range behavior. Several recently proposed functionals
(such as the empirical long-range correction\cite{BSVO04}, the
non-empirical bootstrap kernel\cite{SDSG11} and the
non-empirical meta-GGA kernel\cite{Nazarov2011}) have the
correct long-range behavior; hence, they are promising choices
for the accurate calculation of excitons in TDDFT.
Due to its discrete
nature, the Casida equation approach is rarely used for
periodic systems. For calculating exciton binding energies,
however, it is preferable over the usual response-function
approach.

\section{Method}

In this section we present the details of how we
apply our TDDFT approach for calculating exciton binding energies.
Within the TDA and using the adiabatic approximation
for the xc kernel, Eq.
\parref{eqn:theory:Casida} becomes
\begin{equation}
\sum_{(mn)}\left[\delta_{im}\delta_{jn}(\epsilon_j-\epsilon_i)+F\Hxc^{(ij)(mn)}\right]\rho^{(mn)}(\omega)=\omega\rho^{(ij)}(\omega).
\label{eqn:theory:TDDFTworking}
\end{equation}
We only consider optical excitations which have no momentum
transfer. Thus only the $\vect{q}=0$ part of the xc kernel is
involved. The xc matrix element in reciprocal space for spin-unpolarized systems is given by
\begin{eqnarray}
F\xc^{(ij\vect{k})(mn\vect{k}')}&=&\frac{2}{V}\sum_{\vect{G}\vect{G}'}f_{\text{xc},\vect{G}\vect{G}'}(\vect{q}=0)\nonumber\\
&\times&\matelem{j\vect{k}}{e^{i\vect{G}\cdot\br}}{i\vect{k}}\matelem{m\vect{k}'}{e^{-i\vect{G}'\cdot\br}}{n\vect{k}'},
\label{eqn:theory:fxcworking}
\end{eqnarray}
where $V$ is the volume of the crystal, $\vect{k}$, $\vect{k}'$
are Bloch wavevectors of orbitals, and $i$, $j$, $m$, $n$ are
band indices. Since the matrix element
$\matelem{j\vect{k}}{e^{i\vect{G}\cdot\br}}{i\vect{k}}$
vanishes as $\vect{G}\to0$, the contributions of the head
($\vect{G}=\vect{G}'=0$) and of the wings ($\vect{G}=0$ or
$\vect{G}'=0$) need to be evaluated analytically. $f\xc$ must
diverge as $q^{-2}$ for the head and as $q^{-1}$ for the wings
for these to have a nonzero contribution. In this work, we use
the ABINIT pseudopotential band structure code for the
Kohn-Sham ground state.\cite{ABINIT} Due to the presence of
pseudopotentials, the above matrix element for $\vect{G}=0$
must be replaced by\cite{BR86}
\begin{equation}
\matelem{j\vect{k}}{e^{i\vect{G}\cdot\br}}{i\vect{k}}\to\frac{\matelem{j\vect{k}}
{\hat{p}-i[\hat{r},V_\text{nl}]}{i\vect{k}}}{\epsilon_{j\vect{k}}-\epsilon_{i\vect{k}}},
\end{equation}
where $\hat{p}$ is the momentum operator, $\hat{r}$ is the
position operator, and $V_\text{nl}$ is the nonlocal part of
the pseudopotential.

\subsection{XC kernels} \label{subsec:XC}
We use the following adiabatic xc kernels in our study: the long-range correction,\cite{BSVO04} the
bootstrap kernel,\cite{SDSG11} and the PGG kernel for
singlet excitons,\cite{PGG96,Lein2000} and the hybrid
kernel by Burke {\em et al.} for triplet excitons.\cite{BPG02}

The long-range correction kernel is defined as
\begin{equation}
f\xc^\text{LRC}(\vect{q},\vect{G},\vect{G}')=-\frac{\alpha}{\abs{\vect{q}+\vect{G}}^2}\delta_{\vect{G}\vect{G}'},
\end{equation}
where $\alpha$ is a material-dependent parameter. In Ref.
\onlinecite{BSVO04}, an empirical formula was proposed for
determining $\alpha$:
\begin{equation}
\alpha=4.615\epsilon_\infty^{-1}-0.213,
\label{eqn:theory:LRCempirical}
\end{equation}
where $\epsilon_\infty$ is the high-frequency dielectric
constant. The purpose of this empirical formula is to reproduce
the continuum spectrum; here, we test its effect on the exciton
binding energy, which would be too small to be resolved in common
response-function calculations.\cite{BSVO04} For comparison, we
will also fit $\alpha$ with respect to the experimental exciton
binding energies.

The PGG kernel is a real-space kernel approximating the exact
exchange kernel. In real space, it is defined as\cite{PGG96,Lein2000}
\begin{equation}
f\xc^\text{PGG}(\br,\br')=-\frac{2\abs{\sum_{i\vect{k}}^\text{occ.}\phi_{i\vect{k}}^*(\br)\phi_{i\vect{k}}(\br')}^2}{\abs{\br-\br'}n(\br)n(\br')},
\label{eqn:theory:PGG}
\end{equation}
where $n$ is the ground-state electronic density. We convert the PGG kernel
into reciprocal space for its use in Eq.
\parref{eqn:theory:fxcworking}. The Kohn-Sham orbitals in Eq.
\parref{eqn:theory:PGG} have the Bloch form,
\begin{equation}
\phi_{i\vect{k}}(\br)=\frac{1}{\sqrt{N_\text{cell}}}u_{i\vect{k}}(\br)e^{i\vect{k}\cdot\br},
\end{equation}
where $N_\text{cell}$ is the number of unit cells in the
crystal, and $u_{i\vect{k}}(\br)$ is the Bloch function.
$f\xc^\text{PGG}$ can then be written as
\begin{equation}
f\xc^\text{PGG}(\br,\br')=-\sum_{i\vect{k}}^\text{occ.}\sum_{m\vect{k}'}^\text{occ.}
\frac{2e^{-i(\vect{k}-\vect{k}')\cdot(\br-\br')}}{\abs{\br-\br'}}\: H_{i\vect{k}m\vect{k}'}(\br,\br'),
\end{equation}
where $H_{i\vect{k}m\vect{k}'}(\br,\br')$ is periodic within
one unit cell and defined as
\begin{equation}
H_{i\vect{k}m\vect{k}'}(\br,\br')=\frac{u_{i\vect{k}}^*(\br)u_{i\vect{k}}(\br')u_{m\vect{k}'}(\br)u_{m\vect{k}'}^*(\br')}
{N_\text{cell}^2n(\br)n(\br')}.
\label{eqn:theory:PGGaux}
\end{equation}
The Fourier transform of $f\xc^\text{PGG}$ yields
\begin{eqnarray}
f\xc^\text{PGG}(\vect{q},\vect{G},\vect{G}')
&=&
-\frac{1}{V}\sum_{i\vect{k}}^\text{occ.}\sum_{m\vect{k}'}^\text{occ.}\sum_{\vect{G}_0}\frac{8\pi}{\abs{\vect{q}-(\vect{k}'-\vect{k})+\vect{G}_0}^2}
\nonumber\\
&&
\times\tilde{H}_{i\vect{k}m\vect{k}'}(\vect{G}-\vect{G}_0,\vect{G}'-\vect{G}_0),
\label{eqn:theory:PGGworking}
\end{eqnarray}
where $\tilde{H}$ is obtained by numerical Fourier transform of
expression
\parref{eqn:theory:PGGaux} within one unit cell. For simplicity,
we ignore the local-field effects and only use the head of the
PGG kernel, which is given by
\begin{equation}
f\xc^\text{PGG}(\vect{q},0,0)=-\frac{8\pi}{V}\sum_{i,m,\vect{k}}^\text{occ.}\frac{\tilde{H}_{i\vect{k}m\vect{k}}(0,0)}{\abs{\vect{q}}^2}.
\end{equation}
The PGG kernel is orbital dependent and involves a sum over all
occupied orbitals. It is not obvious whether the
pseudopotential formalism is directly compatible with this
kernel since it does not properly include the core states; this
will be the subject of future study. For the time being, we
use all the occupied pseudo-bands for constructing the PGG
kernel.

The so-called bootstrap kernel\cite{SDSG11} is a recently proposed non-empirical
adiabatic xc kernel, designed to be able to treat excitons:
\begin{equation}
f_{\rm \xcs,sym}^\text{bootstrap}(\vect{q},\vect{G},\vect{G}')=\frac{\epsilon^{-1}_\text{sym}
(\vect{q},\vect{G},\vect{G}')}{\chi_{s,\text{sym}}(\vect{q},\vect{G}=\vect{G}'=0)},
\label{eqn:theory:bootstrap}
\end{equation}
where $f_{\rm \xcs,sym}$, $\epsilon^{-1}_\text{sym}$ and
$\chi_{s,\text{sym}}$ are the xc kernel, inverse dielectric
function and Kohn-Sham linear response function in their
symmetric forms, respectively, defined as
\begin{eqnarray}
f_{\rm \xcs,sym}(\vect{q},\vect{G},\vect{G}')&=&v^{-1/2}_\vect{G}(\vect{q})f\xc(\vect{q},\vect{G},\vect{G}')v^{-1/2}_\vect{G'}(\vect{q}),\\
\epsilon^{-1}_\text{sym}(\vect{q},\vect{G},\vect{G}')&=&v^{-1/2}_\vect{G}(\vect{q})\epsilon^{-1}(\vect{q},\vect{G},\vect{G}')
v^{1/2}_\vect{G'}(\vect{q}),\\
\chi_\text{sym}(\vect{q},\vect{G},\vect{G}')&=&v^{1/2}_\vect{G}(\vect{q})\chi(\vect{q},\vect{G},\vect{G}')v^{1/2}_\vect{G'}(\vect{q}),
\end{eqnarray}
where $v_\vect{G}(\vect{q})=4\pi/\abs{\vect{q}+\vect{G}}^2$ is
the Coulomb potential. $\epsilon^{-1}_\text{sym}$ in Eq.
\parref{eqn:theory:bootstrap} is calculated as
\begin{equation}
\begin{split}
\epsilon^{-1}_\text{sym}&=1+\chi_\text{sym}\\
&=1+(1-\chi_{s,\text{sym}}f_{\rm {\sss HXC},sym})^{-1}\chi_{s,\text{sym}},
\end{split}
\label{eqn:theory:invdielecsym}
\end{equation}
where all quantities in Eq. \parref{eqn:theory:invdielecsym}
are matrices in $\vect{G}$ and $\vect{G}'$. Equations
\parref{eqn:theory:invdielecsym} and
\parref{eqn:theory:bootstrap} are iteratively evaluated until
self-consistency is achieved. Since the head and the wings of
$f\xc$ diverge as $\vect{q}\to0$ (which is important for
excitons), $f\xc$ cannot be used directly in the iteration due
to numerical difficulties. The symmetric forms ensure that no
troubling singularities are involved.

So far, we have neglected the spin. In principle, a
noncollinear spin formulation for the xc
kernel\cite{U12,Wang2004} is needed to treat singlet and
triplet excitations on the same ground. For spin-unpolarized
systems, however, one can define singlet and triplet xc kernels
as
\begin{equation}
f\xc^\text{singlet}=\frac{f\xc^{\uparrow\uparrow}+f\xc^{\uparrow\downarrow}}{2},\quad
f\xc^\text{triplet}=\frac{f\xc^{\uparrow\uparrow}-f\xc^{\uparrow\downarrow}}{2},
\end{equation}
where $f\xc^{\sigma\sigma'}=\delta v_{\text{xc}\sigma}/\delta
n_{\sigma'}$ is the spin-dependent xc kernel. Only $f\xc^{\uparrow\uparrow}$ and $f\xc^{\uparrow\downarrow}$
are involved because $f\xc^{\downarrow\downarrow}=f\xc^{\uparrow\uparrow}$ and $f\xc^{\downarrow\uparrow}=f\xc^{\uparrow\downarrow}$ in spin-unpolarized systems. One calculates
singlet and triplet excitations by performing two separate TDDFT
calculations with $f\xc^\text{singlet}$ and
$f\xc^\text{triplet}$(the triplet calculation does not include
the Hartree kernel). In the spin-dependent case, the important
property for excitons is the exchange splitting $\Delta\x$,
defined as
\begin{equation}
\Delta\x=E_b^\text{triplet}-E_b^\text{singlet}.
\end{equation}

The bootstrap xc kernel was originally developed for the
spin-independent case. Since it is not defined via a functional
derivative, there is no unique way to make it spin dependent.
We propose a plausible spin-dependent generalization of the
bootstrap kernel in such a way that the singlet result does not
change during the self-consistent procedure. The idea is to
replace
$\chi_{s,\text{sym}}^{-1}(\vect{q},\vect{G}=\vect{G}'=0)$ in
Eq.
\parref{eqn:theory:bootstrap} by a matrix $M_{\sigma\sigma'}$:
\begin{eqnarray}
f_{\rm \xcs,sym,\sigma\sigma'}^{\rm bootstrap}(\vect{q},\vect{G},\vect{G}')
&=&
\sum_{\sigma''\vect{G}''}\epsilon^{-1}_{\text{sym},\sigma\sigma''}(\vect{q},\vect{G},\vect{G}'')
\nonumber\\
&& \times M_{\sigma''\sigma'}(\vect{q},\vect{G}'',\vect{G}'),
\label{eqn:theory:generalizedbootstrap}
\end{eqnarray}
where the spin-dependent inverse dielectric function
$\epsilon^{-1}$ for a spin-unpolarized system can be defined as
\begin{equation}
\epsilon^{-1}_{\sigma\sigma'}=\delta_{\sigma\sigma'}+\sum_{\sigma''}v_{\sigma\sigma''}\chi_{\sigma''\sigma'}.
\label{eqn:method:bootstrap:spindep:epsinv}
\end{equation}
The Coulomb interaction is spin-independent
($v_{\sigma\sigma''}=v$), so $\epsilon^{-1}_\text{sym}$ in Eq.
\parref{eqn:method:bootstrap:spindep:epsinv} becomes
\begin{equation}
\epsilon_{\text{sym},\sigma\sigma'}^{-1}=\delta_{\sigma\sigma'}+\chi_\text{sym}^{\uparrow\uparrow}+\chi_\text{sym}^{\uparrow\downarrow}.
\end{equation}
To ensure that $f\xc^\text{singlet}=f\xc^\text{bootstrap}$ during and
after self-consistency, it is straightforward to show that the matrix $M_{\sigma\sigma'}$ must satisfy
the following conditions:
\begin{eqnarray}
M_{\sigma\ne\sigma'}(\vect{q},\vect{G}=\vect{G}')
&=&
-M_{\sigma=\sigma'}(\vect{q},\vect{G}=\vect{G}')
\nonumber\\&&{}
+ 2\chi_{s,\text{sym}}^{-1}(\vect{q},\vect{G}=\vect{G}'=0),\label{eqn:method:Mdiag}\\
M_{\sigma\ne\sigma'}(\vect{q},\vect{G}\ne\vect{G}')
&=&
-M_{\sigma=\sigma'}(\vect{q},\vect{G}\ne\vect{G}').\label{eqn:method:Moffdiag}
\end{eqnarray}
This ensures that the singlet xc kernel reproduces the bootstrap kernel, but it also has
consequences for the triplet xc kernel. If the $M_{\sigma\sigma'}$ matrix is the same for
all steps of the iteration, it is easy to show that during the iteration
$f\xc^\text{triplet}$ would not change, and it can be determined without
any iterative procedure as
\begin{equation}
f_{\xcs,\text{sym}}^\text{triplet}(\vect{q},\vect{G},\vect{G}')=\frac{1}{2} \left[M_{\uparrow\uparrow}(\vect{q},\vect{G},\vect{G}')-
M_{\uparrow\downarrow}(\vect{q},\vect{G},\vect{G}')\right].
\end{equation}
$f\xc^\text{triplet}$ is determined self-consistently only
if $M_{\sigma\sigma'}$ depends on $f\xc$ of the previous iteration step.


Burke, Petersilka, and Gross\cite{BPG02} proposed a hybrid
spin-dependent xc kernel, where
\begin{eqnarray}
f^{\rm hybrid}_{\rm \sss XC, \uparrow\uparrow}&=& f^{\rm PGG}_{\rm \sss XC, \uparrow\uparrow},\\
f^{\rm hybrid}_{\rm \sss XC, \uparrow\downarrow}&=&f^{\rm ALDA}_{\rm \sss XC, \uparrow\downarrow}.
\end{eqnarray}
This hybrid kernel yields rather accurate singlet-triplet splittings as
well as excitations frequencies in finite system. Inspired by this, we propose to calculate
the singlet-triplet splitting of excitons with a similar hybrid kernel. Since the bootstrap kernel is a
parameter-free xc kernel that has good performance for exciton
binding energies in several tested systems, we use the
bootstrap kernel as the singlet xc kernel:
\begin{eqnarray}\label{eqn:theory:ourhybrid1}
f^{\rm bootstrap}_{\xcs, \uparrow\uparrow}&=& 2f\xc^{\text{bootstrap}}-
f^{\rm ALDA}_{\xcs,\uparrow\downarrow},\\
f^{\rm bootstrap}_{\xcs,\uparrow\downarrow}&=&f^{\rm ALDA}_{\xcs, \uparrow\downarrow}.
\label{eqn:theory:ourhybrid2}
\end{eqnarray}
The singlet exciton binding energy retains the result of the
bootstrap kernel. This modified hybrid kernel is effectively an instance
of the generalized spin-dependent bootstrap kernel, where
\begin{equation}
M_{\uparrow\uparrow}=f_{\xcs,\text{sym}}^\text{bootstrap}-f_{\xcs,\text{sym},\uparrow\downarrow}^\text{ALDA}+2\chi_{s,\text{sym}}^{-1},
\end{equation}
and $M_{\uparrow\downarrow}$ is determined by Eqs. \parref{eqn:method:Mdiag} and \parref{eqn:method:Moffdiag}.

\subsection{Computational aspects and numerical details}

Let us now briefly summarize the computational aspects of our exciton calculations;
more technical details will be presented elsewhere.\cite{code}

An LDA ground-state calculation is carried out with the ABINIT
pseudopotential code.\cite{ABINIT} The ground-state band
structure and Bloch functions are taken as the input to our
TDDFT calculations. Due to the absence of the derivative
discontinuity, the LDA Kohn-Sham gap is too small to
approximate the quasiparticle gap.\cite{Perdew1983} This
introduces big errors in the corresponding TDDFT calculation,
since adiabatic xc kernels cannot change the
gap.\cite{Gonze1999} A frequency-dependent kernel together with a good ground-state xc functional would be
required to fully treat the gap problem in TDDFT. However,
since our focus is on the exciton binding energies and not on
the gap itself, we shift the gap to its corresponding
experimental value by applying a simple scissor operator
\cite{Levine1989} to the conduction bands, and we apply a
corresponding correction to the momentum matrix elements.\cite{DG93} For xc kernels that explicitly
depend on the density, we use the pseudodensity with the
correction described in Ref. \onlinecite{T93}.

The $F\xc$ matrix in Eq. \parref{eqn:theory:fxcworking} is
represented in the transition space of Kohn-Sham excitations. It has the dimension
$N_\text{v}\times N_\text{c}\times N_\vect{k}$, where v stands
for valence bands, c stands for conduction bands, and
$N_\vect{k}$ is the number of $\vect{k}$-grid points in the
Brillouin zone. To achieve convergence, the dimension must be
large, so a Casida-equation-type calculation is computationally
more demanding than the usual response-function calculation; in
particular, storage of a large matrix is required. For this reason, Casida-equation-type
calculations are not usually done for periodic systems. In
contrast, a response-function calculation processes matrices in
reciprocal space; the dimension of the matrix equals the number
of reciprocal lattice vectors used in the calculation, which is
much smaller than the size of the $F\xc$ matrix [Eq.
(\ref{eqn:theory:fxcworking})] required for convergence.

According to the Wannier model, excitons are dominated by
single-particle (Kohn-Sham) excitations near the band edge for
direct-gap solids. Although strictly speaking the Wannier model
refers to quasiparticle bands instead of Kohn-Sham bands, the
Kohn-Sham band structure is similar to the quasiparticle band
structure near the Fermi level (aside from having the wrong
gap).\cite{BSVO04} Thus for direct-gap solids, only the highest
valence bands and the lowest conduction bands are needed in the
calculation (including degeneracies at the $\Gamma$ point).

With only a few bands used, the dimension of the matrices in Eq.
\parref{eqn:theory:Casida} is vastly reduced, so the eigenvalue problem can be
solved with acceptable cost. Although this
restriction to a few bands near the band gap is sufficient for
describing excitons, this would yield unsatisfactory continuum
spectra; but those are not our concern in this work, since they
can be calculated much better with the standard response-function
approach of TDDFT.

To make the numerical problem more manageable for
resource-limited environments, the storage of the large
matrices can be distributed among different nodes of a computer
cluster, and we use the ScaLAPACK\cite{ScaLAPACK} package to
manipulate the distributed matrix. Since we only need the first
exciton binding energy, we can further simplify the eigenvalue
problem by using an iterative eigenvalue solver such as
FEAST\cite{Polizzi2009} to converge only the eigenvalue for the
exciton instead of all the excitations. Iterative eigenvalue
solvers require a predefined range of the desired eigenvalues,
which would require knowledge about the result before it is
obtained. But in our case the range of frequency is
conveniently chosen from zero to the band gap, since excitonic
excitations are always below the band gap. With all the
techniques described in this section, Casida-equation-like
calculations for periodic solids become practically feasible.

We check convergence against the following parameters:
the number of bands in the calculation, the
$\vect{k}$-grid for the ground state, and the reciprocal lattice
vectors in constructing the matrix in Eq.
\parref{eqn:theory:TDDFTworking}.

For all systems under consideration, the highest valence band
has $p$-character and is triply degenerate at the $\Gamma$
point, while there is no degeneracy at the $\Gamma$ point for
the lowest conduction band. Thus we only include 3 valence
bands and 1 conduction band in Eq.
\parref{eqn:theory:TDDFTworking}. We tested the convergence with
respect to number of bands, and we found that the effect in the
exciton binding energy from increasing the number of valence
bands is much smaller than that of increasing the number of
$\vect{G}$-vectors. We test convergence
against the TDDFT response-function calculations in Ref. \onlinecite{SDSG11}, which by nature include
much more bands than we use here. The excitons are strong enough
in Ar and LiF to be resolved with the response-function approach, and our calculation with
three valence and one conduction bands produces exciton binding energies
very close to those reported in Ref. \onlinecite{SDSG11} (see Table \ref{table1}).
We thus conclude that our few-band
approach is sufficient for exciton binding energies.

We find that for strongly bound Frenkel excitons in
insulators, a small $\vect{k}$-grid such as a
$10\times10\times10$ Monkhorst-Pack grid\cite{Monkhorst1976} is
sufficient for the convergence of the binding energy. In the
case of weakly bound Wannier excitons in semiconductors, the
convergence is much slower and requires at least a
$18\times18\times18$ grid. This result is not surprising, since
Frenkel excitons are local in real space, which means they are
diffuse in reciprocal space, and therefore require less
resolution to be well-described in the reciprocal space than
Wannier excitons.

Several previous TDDFT studies (within the response-function
approach) state that convergence is achieved for semiconductors
with grids like $15\times15\times15$.\cite{BSVO04,SDSG11}
However, in these cases convergence is to be understood with
respect to the continuum spectra instead of the exciton binding
energies. We find that with a $15\times15\times15$ grid, the
maximum relative error in exciton binding energies for
semiconductors studied in this work is $139\%$. We therefore
calculate zincblende materials with a $18\times18\times18$
grid, wurtzite materials with a $20\times20\times20$ grid, and
insulators with a $10\times10\times10$ grid. The ground-state
calculation with a bigger $\vect{k}$-grid is feasible since the
number of effective $\vect{k}$-points can be greatly reduced by
symmetry, but this is not the case for TDDFT, where the
$\vect{k}$-points in the entire first Brillouin zone must be
included.

Excitons are usually described as electron-hole pairs and are
modeled with parabolic bands. This may suggest that one can get
away with using only $\vect{k}$-points near the $\Gamma$-point
($\vect{k}=0$), which would greatly decrease the size of the
problem. This approach was employed by Rohlfing and Louie in a
BSE study.\cite{RL98} In a TDDFT context, however,
 for GaAs including $62\%$ of the $\vect{k}$-points (centered at the $\Gamma$-point)
induces a $74\%$ relative error, and even including $95\%$ of
the $\vect{k}$-points still induces a $5\%$ error, and the
benefit of calculation speed is diminishing. We confirm the
result with BSE and TDDFT calculations for an 1D Kronig-Penney
model,\cite{YLU12} where we find that only including $k$-points
near $k=0$ induces a relative error three times larger in TDDFT
than in BSE. This can be understood by analyzing the coupling
matrix of TDDFT [$F\xc$ in Eq.
\parref{eqn:theory:fxcworking}] and BSE (Ref.
\onlinecite{YLU12}): the equivalent object of $F\xc$ in BSE is
dominated by its diagonal ($\vect{k}=\vect{k}'$) part, so only taking the
$\vect{k}$-points near the $\Gamma$-point still retains its
shape. In TDDFT this is not the case. Excitons are known to
have collective character, but this discrepancy between model
systems and excitons in real materials points to a surprisingly
high degree of collectiveness when represented with
single-particle Kohn-Sham excitations, since Kohn-Sham
excitations contribute over the entire Brillouin zone.

The head of the xc kernel makes the
largest contribution in Eq. \parref{eqn:theory:TDDFTworking};
the other contributions can be usually ignored without loss of
accuracy. We find that this is
also true for the exciton binding energy. We checked the error
in the exciton binding energy introduced by only using the head
of the xc kernel, and the error is about $1\%$ in most cases, and
less than $5\%$ throughout. Thus we only use the head
in most calculations, but in cases where the head vanishes
(such as the ALDA part included in the hybrid kernel), we
include $\vect{G}$-vectors with length up to $2G_0$, with
$G_0$ being the longest reciprocal cell vector. The
construction of the bootstrap kernel involves matrix operations
in the reciprocal space, so we also use $\vect{G}$-vectors up
to $2G_0$ for it. After the bootstrap kernel is calculated, we
only use its head in Eq. \parref{eqn:theory:TDDFTworking}. It
should be noticed that although this still yields acceptable
accuracy for the exciton binding energy, only using the head
would not produce more than one exciton.\cite{YLU12} If one
needs an excitonic Rydberg series, the wings and body
contribution of the xc kernel are necessary.

For the bootstrap kernel there is an additional convergence
issue related to the number of bands that need to be included
in the Kohn-Sham response function $\chi_s$. For large-gap
insulators, we only need to use four bands in the calculation
of $\chi_s$; For zincblende semiconductors, we have to use a
total of 60 bands to achieve convergence. For wurtzite
semiconductors, we use 10 valence bands and 40 conduction bands
in the calculation.

\begin{table*}
\caption{Lowest singlet exciton binding energies $E_b^\text{singlet}$ and singlet-triplet splittings $\Delta\x$, 
calculated with TDDFT using various different xc kernels, and compared to experimental values from the literature.
A star ($*$) means that no bound exciton was obtained from the calculation, ``n.c.'' means that no calculation was performed.} \label{table1}
\begin{tabular}{l|cccccccc}\hline
Material\footnotemark[1] & GaAs & $\beta$-GaN & $\alpha$-GaN & CdS & CdSe & Ar & Ne & LiF\\
\hline\hline
Exp. gap (eV) & 1.52 & 3.3 & 3.452 & 2.42 & 1.74 & 14.25 & 21.51 & 14.20\\
Exp. $E_b^\text{singlet}$\footnotemark[2] & 3.27meV & 26.0meV & 20.4meV & 28.0meV & 15.0meV & 1.90eV & 4.08eV & 1.6eV\\
Exp. $\Delta\x$\footnotemark[3] & 9.61$\mu$eV & --- & --- &--- & 49.78$\mu$eV & 0.16eV & 0.14eV & --- \\
\hline
LRC empirical $\alpha$\footnotemark[4] & 0.211 & 0.6578 & 0.6496 & 0.6448 & 0.4721 & 2.697 & --- & 2.191\\
LRC empirical $E_b^\text{singlet}$ & 0.8580meV  & 0.5143meV  & $*$  & 0.5131meV & 1.405meV  & 0.3043meV & --- & 1.136meV\\
LRC fit $\alpha$ & 0.595 & 2.409 & 3.6285 & 4.244 & 2.144  & 21.45 & 96.5 & 9.5 \\
\hline
$f\xc^\text{bootstrap}$ $E_b^\text{singlet}$ & 0.3318meV & 0.1992meV & $*$ & 0.4610meV & 0.8947meV & 2.156eV & 6.225eV & 1.547eV\\
Corresponding LRC $\alpha$\footnotemark[5] & 0.08836 & 0.3048 & 0.2147 & 0.5895 & 0.3183 & 22.6324 & 126.673 & 9.32326\\
\hline
PGG $E_b^\text{singlet}$ & $*$ & n.c. & n.c. & n.c. & n.c. & $*$ & n.c. & $*$ \\
Corresponding LRC $\alpha$\footnotemark[5] & $5.820\times10^{-5}$ & n.c. & n.c. & n.c. & n.c. & $3.924\times10^{-4}$ & n.c. & $3.851\times10^{-4}$\\
\hline
hybrid\footnotemark[6] $\Delta\x$ & 37.65$\mu$eV & 40.51$\mu$eV & $*$ & n.c. & n.c. & 0.03297eV & 0.01287eV & 0.5041meV
\\ \hline
\end{tabular}\\
\footnotetext[1]{Unless otherwise mentioned, zincblende materials are
calculated with $18\times18\times18$ Monkhorst-Pack $\vect{k}$-point
grid,  wurtzite materials are calculated with $20\times20\times20$ grid, and solid Ar, solid Ne, and LiF are calculated with
$10\times10\times10$ grid.}
\footnotetext[2]{Experimental data from Refs. \onlinecite{PCK92,ASWS97,MLSK97,JSKG94,VSS79,HKKS69,RW67,SK79}. }
\footnotetext[3]{Experimental data from Refs. \onlinecite{ELB79,KRU75,HKKS69,SK79}. No experimental data available for
GaN, CdS, and LiF.}
\footnotetext[4]{Calculated using Eq. \parref{eqn:theory:LRCempirical}. The $\epsilon_\infty^{-1}$ data is not available for solid Ne.}
\footnotetext[5]{The head of the xc kernel has the same form as the LRC.}
\footnotetext[6]{See Eqs. \parref{eqn:theory:ourhybrid1} and \parref{eqn:theory:ourhybrid2}.}
\end{table*}

\section{Results}
\label{sect:results}

We consider several common direct-gap zincblende (GaAs,
$\beta$-GaN) and wurtzite ($\alpha$-GaN, CdS, CdSe)
semiconductors, as well as insulators (LiF, solid Ar, solid
Ne). Results are presented in Table \ref{table1}. The first
three rows give experimental data on the band
gap,\cite{GSS87,RNLP94,MLSK97,SB05,GP05,PLO76} the binding
energy $E_b^{\rm singlet}$ of the lowest singlet
exciton,\cite{PCK92,ASWS97,MLSK97,JSKG94,VSS79,HKKS69,RW67,SK79}
and the singlet-triplet exchange splitting $\Delta\x$.\cite{ELB79,KRU75,HKKS69,SK79} The remaining rows of Table
\ref{table1} show the results of our calculations.

The exciton binding energies calculated with the LRC kernel are generally
found to be significantly too small if the empirical $\alpha$ is used, see Eq. \parref{eqn:theory:LRCempirical};
for $\alpha$-GaN, there is not even a bound exciton.
We therefore determine the LRC $\alpha$ parameter for the exciton binding
energies by fitting to experimental data for $E_b^{\rm singlet}$, and we find that they are quite different from the
empirical formula for $\alpha$.
Ref. \onlinecite{BSVO04} argued that $\alpha$ must be
proportional to $\epsilon_\infty^{-1}$, but we cannot confirm
such a linear fit for $\alpha$ with our results.
The sensitivity of the exciton binding energy with respect to
$\alpha$ varies a lot for different materials. Though the
empirical formula for $\alpha$ [Eq. \parref{eqn:theory:LRCempirical}] was originally not developed to give
accurate exciton binding energies, we find that calculations
with empirical $\alpha$ still yield bound excitons (except for $\alpha$-GaN). This may
explain, at least in part, why these parameters lead to quite accurate spectra in the vicinity of the gap.

We  next consider two nonempirical kernels, bootstrap and PGG. We
find that the self-consistent procedure of the bootstrap kernel
is very stable: even if we start the iteration with a different
$f\xc$ (instead of starting with no $f\xc$),
we always converge to the same bootstrap
kernel. We confirm that the bootstrap kernel produces accurate
exciton binding energies for Frenkel excitons in Ar and LiF, as
reported in Ref. \onlinecite{SDSG11}. For Ne the
bootstrap kernel overbinds by about 50\%, but it still yields the correct
order of magnitude for $E_b^{\rm singlet}$. For Wannier
excitons in the studied semiconductors, however, the bootstrap kernel
fails to differentiate between different materials and in all cases yields
exciton binding energies that are too low. On the other hand,
the excitonic enhancement of the
continuum spectrum is reported to be well-described by this kernel.
This is understandable since the corresponding LRC $\alpha$
parameters for semiconductors are close to those given by
the empirical formula Eq. \parref{eqn:theory:LRCempirical}.

The performance of the PGG kernel, which works well in finite
systems, is disappointing: it does not produce any bound excitons at all,
despite having a nonzero head contribution. The
PGG kernel is an exchange-only kernel, and is known to
bind quite strongly in finite systems,\cite{PGG96,BPG02} so it is surprising that it does
not yield any bound exciton in the cases we tested. One 
possible reason is that the pseudopotential treatment is not compatible
with the explicit orbital dependence in the PGG kernel, since the contribution from core
orbitals cannot be systematically included.
Also it should be noted that while periodic systems are
dominated by the head of the xc kernel,
there is no corresponding effect in finite systems.
This is because in finite systems the electron dynamics can
be viewed as coming entirely from local-field effects.
Thus, the strongly attractive nature of the PGG kernel
in finite systems would at most translate into a strong body of
the xc matrix in periodic systems (which, however, is irrelevant for excitons), but does not
necessarily guarantee a strong head. This is indeed confirmed by calculating
the LRC $\alpha$ that corresponds to the PGG kernel, which turns out to be
orders of magnitude too weak (see the second-to-last row of Table \ref{table1}).

As we discussed  in Section \ref{subsec:XC}, not many long-range spin-dependent xc kernels
for treating triplet excitons are known. Our hybrid kernel [Eqs. \parref{eqn:theory:ourhybrid1} and \parref{eqn:theory:ourhybrid2}]
can be viewed as a special case of
the generalized bootstrap kernel in Eq.
\parref{eqn:theory:generalizedbootstrap}, which performed well for singlet excitons.
We find that the exchange splitting is of the right order of magnitude
for GaAs, although somewhat too large. A similar $\Delta\x$ is found for $\beta$-GaN, but there is no
experimental data available for comparison. Since $\alpha$-GaN does not have a bound exciton with the
bootstrap kernel, it is not possible to obtain a well-defined $\Delta\x$ with the hybrid kernel.
No calculation was performed for CdS and CdSe due to excessive memory requirements. Finally,
for the strongly bound excitons in the large-gap
insulators Ar, Ne, and LiF,  $\Delta\x$ comes out significantly too small.

%
%

Considering the fact that the singlet-triplet exchange splitting is essentially treated
on an ALDA level, it is perhaps not surprising that the results for $\Delta\x$ with
the hybrid kernel are not terribly accurate. Clearly, more sophisticated spin-dependent
xc kernels for singlet and triplet excitons need to be developed.

\section{Conclusion}
\label{sect:conclusion}

In this paper we have introduced an alternative TDDFT approach for calculating
excitonic binding energies in solids. We present the first converged
Casida-equation-type TDDFT calculations for several
materials, showing that such calculations are feasible for real
periodic bulk systems. The approach yields exciton binding energies directly,
rather than the optical spectrum. Although using only a few bands
in general does not yield accurate continuum spectra, it is
sufficient for the convergence of exciton binding energies.
Binding energies of Frenkel excitons converge quickly with
respect to the $\vect{k}$-grid, while Wannier excitons require
larger $\vect{k}$-grids than usually seen in the literature.
Although excitons are conventionally described as bound electron-hole pairs,
only taking $\vect{k}$-points near the $\Gamma$-point does not
give a good description for excitons in TDDFT, suggesting very strong
collective character when represented with Kohn-Sham excitations.

We test our formalism with several xc kernels. The LRC
empirical formula, whose empirical parameter has been designed
for reproducing the continuum spectrum, usually produces bound
excitons as well, though the binding energies are generally too
small. This is of course hardly surprising, because it is
difficult to imagine how a single parameter could be sufficient
to fit all aspects of the optical response. If one is
interested in bound excitons rather than the continuum
spectrum, the strength of the LRC kernel has to be increased.

The bootstrap kernel is generally accurate for Frenkel
excitons, while it produces Wannier excitons that are somewhat
too weakly bound. On the other hand, the PGG kernel does not
yield any bound excitons at all. Thus, at present we do not know of
any simple, nonempirical xc kernel that produces accurate
bound Wannier excitons in solids. xc kernels derived from many-body
theory\cite{Reining2002,ORR02,Sottile2003,Marini2003} may be
expected to perform better than the kernels we have studied here, but they are significantly more complex.

We also extended our formalism to triplet excitons, and derived
a formula to generalize the bootstrap kernel to spin-dependent
systems, with the hybrid kernel as a special case. This hybrid
kernel yields the correct order of magnitude for
singlet-triplet exchange splitting in some cases, but is in general not very quantitatively
accurate. Thus, we have given a proof of principle that TDDFT is capable of producing reasonable
exchange splittings in solids;  the search for more accurate spin-dependent xc kernels
for triplet excitons remains an important task for future investigations.

In summary, we have shown that TDDFT shows considerable promise for treating excitonic effects,
but more accurate multipurpose xc kernels
for solids are needed, particularly for spin-dependent phenomena. Our
approach for directly calculating exciton binding energies will be convenient for
facilitating such future developments.

\section*{Acknowledgement}
We thank Sangeeta Sharma and Hardy Gross for helpful discussions.
This work was supported by the National Science Foundation
Grant No. DMR-1005651.

\bibliography{draft}

\begin{thebibliography}{10}

\bibitem{Koch2006}
S.~W. Koch, M.~Kira, G.~Khitrova, and H.~M. Gibbs.
\newblock {\em Nature Mat.}, 5:523, 2006.

\bibitem{Scholes2006}
G.~D. Scholes and G.~Rumbles.
\newblock {\em Nature Mat.}, 5:683, 2006.

\bibitem{HK09}
H.~Haug and S.~W. Koch.
\newblock {\em Quantum theory of the optical and electronic properties of
  semiconductors}.
\newblock World scientific, 5th edition, 2009.

\bibitem{W37}
G.~H. Wannier.
\newblock {\em Phys. Rev.}, 52:191, 1937.

\bibitem{D56}
G.~Dresselhaus.
\newblock {\em J. Phys. Chem. Solids}, 1:14, 1956.

\bibitem{ORR02}
G.~Onida, L.~Reining, and A.~Rubio.
\newblock {\em Rev. Mod. Phys.}, 74:601, 2002.

\bibitem{Puschnig2002}
P.~Puschnig and C.~Ambrosch-Draxl.
\newblock {\em Phys. Rev. B}, 66:165105, 2002.

\bibitem{Fuchs2008}
F.~Fuchs, C.~R{\"o}dl, A.~Schleife, and F.~Bechstedt.
\newblock {\em Phys. Rev. B}, 78:085103, 2008.

\bibitem{Ramos2008}
L.~E. Ramos, J.~Paier, G.~Kresse, and F.~Bechstedt.
\newblock {\em Phys. Rev. B}, 78:195423, 2008.

\bibitem{Setten2011}
M.~J. {van Setten}, R.~Gremaud, G.~Brocks, B.~Dam, G.~Kresse, and G.~A. {de
  Wijs}.
\newblock {\em Phys. Rev. B}, 83:035422, 2011.

\bibitem{MMNG12}
M.~A.~L. Marques, N.~T. Maitra, F.~M.~S. Nogueira, E.~K.~U. Gross, and
  A.~Rubio, editors.
\newblock {\em Fundamentals of time-dependent density functional theory}.
\newblock Lecture notes in physics. Springer, Berlin, 2012.

\bibitem{U12}
C.~A. Ullrich.
\newblock {\em Time-dependent density-functional theory: concepts and
  applications}.
\newblock Oxford University Press, Oxford, 2012.

\bibitem{Botti2007}
S.~Botti, A.~Schindlmayr, R.~{Del Sole}, and L.~Reining.
\newblock {\em Rep. Prog. Phys.}, 70:357, 2007.

\bibitem{Yabana2012}
K.~Yabana, T.~Sugiyama, Y.~Shinohara, T.~Otobe, and G.~F. Bertsch.
\newblock {\em Phys. Rev. B}, 85:045134, 2012.

\bibitem{Reining2002}
L.~Reining, V.~Olevano, A.~Rubio, and G.~Onida.
\newblock {\em Phys. Rev. Lett.}, 88:066404, 2002.

\bibitem{Kim2002}
Y.-H. Kim and A.~G{\"o}rling.
\newblock {\em Phys. Rev. Lett.}, 89:096402, 2002.

\bibitem{Sottile2003}
F.~Sottile, V.~Olevano, and L.~Reining.
\newblock {\em Phys. Rev. Lett.}, 91:056402, 2003.

\bibitem{Marini2003}
A.~Marini, R.~{Del Sole}, and A.~Rubio.
\newblock {\em Phys. Rev. Lett.}, 91:256402, 2003.

\bibitem{Nazarov2011}
V.~U. Nazarov and G.~Vignale.
\newblock {\em Phys. Rev. Lett.}, 107:216402, 2011.

\bibitem{GGG97}
Ph. Ghosez, X.~Gonze, and R.~W. Godby.
\newblock {\em Phys. Rev. B}, 56:12811, 1997.

\bibitem{BSVO04}
S.~Botti, F.~Sottile, N.~Vast, V.~Olevano, L.~Reining, H.-C. Weissker,
  A.~Rubio, G.~Onida, R.~Del~Sole, and R.~W. Godby.
\newblock {\em Phys. Rev. B}, 69:155112, 2004.

\bibitem{Bruneval2006}
F.~Bruneval, F.~Sottile, V.~Olevano, and L.~Reining.
\newblock {\em J. Chem. Phys.}, 124:144113, 2006.

\bibitem{Sottile2007}
F.~Sottile, M.~Marsili, V.~Olevano, and L.~Reining.
\newblock {\em Phys. Rev. B}, 76:161103, 2007.

\bibitem{TU08}
V.~Turkowski and C.~A. Ullrich.
\newblock {\em Phys. Rev. B}, 77:075204, 2008.

\bibitem{TLU09}
V.~Turkowski, A.~Leonardo, and C.~A. Ullrich.
\newblock {\em Phys. Rev. B}, 79:233201, 2009.

\bibitem{YLU12}
Z.-H. Yang, Y.~Li, and C.~A. Ullrich.
\newblock {\em J. Chem. Phys.}, 137:014513, 2012.

\bibitem{C96}
M.~E. Casida.
\newblock Time-dependent density functional response theory of molecular
  systems: theory, computational methods, and functionals.
\newblock In J.~M. Seminario, editor, {\em Recent developments and applications
  in density functional theory}. Elsevier, Amsterdam, 1996.

\bibitem{SDSG11}
S.~Sharma, J.~K. Dewhurst, A.~Sanna, and E.~K.~U. Gross.
\newblock {\em Phys. Rev. Lett.}, 107:186401, 2011.

\bibitem{Ulbrich1985}
R.~G. Ulbrich.
\newblock {\em Adv. Solid State Phys.}, 25:299, 1985.

\bibitem{Uihlein1981}
C.~Uihlein, D.~Fr{\"o}hlich, and R.~Kenklies.
\newblock {\em Phys. Rev. B}, 23:2731, 1981.

\bibitem{SR66}
L.~J. Sham and T.~M. Rice.
\newblock {\em Phys. Rev.}, 144:708, 1966.

\bibitem{Yabana2006}
K.~Yabana, T.~Nakatsukasa, J.-I. Iwata, and G.~F. Bertsch.
\newblock {\em Phys. Stat. Sol. (b)}, 243:1121, 2006.

\bibitem{GiulianiVignale}
G.~F. Giuliani and G.~Vignale.
\newblock {\em Quantum Theory of the Electron Liquid}.
\newblock Cambridge University Press, Cambridge, 2005.

\bibitem{Gross1985}
E.~K.~U. Gross and W.~Kohn.
\newblock {\em Phys. Rev. Lett.}, 55:2850, 1985.
\newblock Erratum: {\em ibid.} {\bf 57}, 923 (1986).

\bibitem{Ullrich2009}
C.~A. Ullrich.
\newblock {\em J. Chem. Theor. Comput.}, 5:859, 2009.

\bibitem{ABINIT}
X.~Gonze, B.~Amadon, P.-M. Anglade, J.-M. Beuken, F.~Bottin, P.~Boulanger,
  F.~Bruneval, D.~Caliste, R.~Caracas, M.~C{\^ot}{\'e}, T.~Deutsch,
  L.~Genovese, Ph. Ghosez, M.~Giantomassi, S.~Goedecker, D.~R. Hamann,
  P.~Hermet, F.~Jollet, G.~Jomard, S.~Leroux, M.~Mancini, S.~Mazevet, M.~J.~T.
  Oliveira, G.~Onida, Y.~Pouillon, T.~Rangel, G.-M. Rignanese, D.~Sangalli,
  R.~Shaltaf, M.~Torrent, M.~J. Verstraete, G.~Zerah, and J.~W. Zwanziger.
\newblock {\em Computer Phys. Comm.}, 180:2582, 2009.

\bibitem{BR86}
S.~Baroni and R.~Resta.
\newblock {\em Phys. Rev. B}, 33:7017, 1986.

\bibitem{PGG96}
M.~Petersilka, U.~J. Gossmann, and E.~K.~U. Gross.
\newblock {\em Phys. Rev. Lett.}, 76:1212, 1996.

\bibitem{Lein2000}
M.~Lein, E.~K.~U. Gross, and J.~P. Perdew.
\newblock {\em Phys. Rev. B}, 61:13431, 2000.

\bibitem{BPG02}
K.~Burke, M.~Petersilka, and E.~K.~U. Gross.
\newblock A hybrid functional for the exchange-correlation kernel in
  time-dependent density functional theory.
\newblock In V.~Barone, P.~Fantucci, and A.~Bencini, editors, {\em Recent
  Advances in Density Functional Methods}, volume III, page~67. World
  Scientific Press, Singapore, 2002.

\bibitem{Wang2004}
F.~Wang and T.~Ziegler.
\newblock {\em J. Chem. Phys.}, 121:12191, 2004.

\bibitem{code}
Z.-H. Yang and C.~A. Ullrich.
\newblock arXiv: 1303.2637.

\bibitem{Perdew1983}
J.~P. Perdew and M.~Levy.
\newblock {\em Phys. Rev. Lett.}, 51:1884, 1983.

\bibitem{Gonze1999}
X.~Gonze and M.~Scheffler.
\newblock {\em Phys. Rev. Lett.}, 82:4416, 1999.

\bibitem{Levine1989}
Z.~H. Levine and D.~C. Allan.
\newblock {\em Phys. Rev. Lett.}, 63:1719, 1989.

\bibitem{DG93}
R.~{Del Sole} and R.~Girlanda.
\newblock {\em Phys. Rev. B}, 48:11789, 1993.

\bibitem{T93}
M.~Teter.
\newblock {\em Phys. Rev. B}, 48:5031, 1993.

\bibitem{ScaLAPACK}
L.~S. Blackford, J.~Choi, A.~Cleary, E.~D'Azevedo, J.~Demmel, I.~Dhillon,
  J.~Dongarra, S.~Hammarling, G.~Henry, A.~Petitet, K.~Stanley, D.~Walker, and
  R.~C. Whaley.
\newblock {\em {ScaLAPACK} Users' Guide}.
\newblock Society for Industrial and Applied Mathematics, Philadelphia, 1997.

\bibitem{Polizzi2009}
E.~Polizzi.
\newblock {\em Phys. Rev. B}, 79:115112, 2009.

\bibitem{Monkhorst1976}
H.~J. Monkhorst and J.~D. Pack.
\newblock {\em Phys. Rev. B}, 13:5188, 1976.

\bibitem{RL98}
M.~Rohlfing and S.~G. Louie.
\newblock {\em Phys. Rev. Lett.}, 81:2312, 1998.

\bibitem{PCK92}
M.~Parenteau, C.~Carlone, and S.~M. Khanna.
\newblock {\em J. Appl. Phys.}, 71:3747, 1992.

\bibitem{ASWS97}
D.~J. As, F.~Schmilgus, C.~Wang, B.~Sch\"{o}ttker, D.~Schikora, and K.~Lischka.
\newblock {\em Appl. Phys. Lett.}, 70:1311, 1997.

\bibitem{MLSK97}
J.~F. Muth, J.~H. Lee, I.~K. Shmagin, R.~M. Kolbas, H.~C. Casey, B.~P. Keller,
  U.~K. Mishra, and S.~P. DenBaars.
\newblock {\em Appl. Phys. Lett.}, 71:2572, 1997.

\bibitem{JSKG94}
M.~A. Jakobson, V.~D. Kagan, R.~P. Seisyan, and E.~V. Goncharova.
\newblock {\em J. Cryst. Growth}, 138:225, 1994.

\bibitem{VSS79}
J.~Voigt, F.~Spiegelberg, and M.~Senoner.
\newblock {\em Phys. Status Solidi B}, 91:189, 1979.

\bibitem{HKKS69}
R.~Haensel, G.~Keitel, E.~E. Koch, M.~Skibowski, and P.~Schreiber.
\newblock {\em Phys. Rev. Lett.}, 23:1160, 1969.

\bibitem{RW67}
D.~M. Roessler and W.~C. Walker.
\newblock {\em J. Opt. Soc. Am.}, 57:835, 1967.

\bibitem{SK79}
V.~Saile and E.~E. Koch.
\newblock {\em Phys. Rev. B}, 20:784, 1979.

\bibitem{ELB79}
W.~Ekardt, K.~L\"{o}sch, and D.~Bimberg.
\newblock {\em Phys. Rev. B}, 20:3303, 1979.

\bibitem{KRU75}
V.~A. Kiselev, B.~S. Razbirin, and I.~N. Uraltsev.
\newblock {\em Phys. Status Solidi B}, 72:161, 1975.

\bibitem{GSS87}
R.~W. Godby, M.~Schl\"{u}ter, and L.~J. Sham.
\newblock {\em Phys. Rev. B}, 35:4170, 1987.

\bibitem{RNLP94}
G.~Ram\'{i}rez-Flores, H.~Navarro-Contreras, A.~Lastras-Mart\'{i}nez, R.~C.
  Powell, and J.~E. Greene.
\newblock {\em Phys. Rev. B}, 50:8433, 1994.

\bibitem{SB05}
B.~Streetman and S.~Banerjee.
\newblock {\em Solid state electronic devices}.
\newblock Prentice Hall, New Jersey, 5th edition, 2005.

\bibitem{GP05}
S.~Galami\'{c}-Mulaomerovi\'{c} and C.~H. Patterson.
\newblock {\em Phys. Rev. B}, 72:035127, 2005.

\bibitem{PLO76}
M.~Piacentini, D.~W. Lynch, and C.~G. Olson.
\newblock {\em Phys. Rev. B}, 13:5530, 1976.

\end{thebibliography}
\bibliographystyle{unsrt}

\end{document}